\documentclass[acmsmall]{acmart}
\usepackage{soul}
\usepackage{xcolor}

\definecolor{lightblue}{RGB}{173, 216, 230} 
\definecolor{lightyellow}{RGB}{255, 255, 153}

\AtBeginDocument{%
  }

\setcopyright{acmlicensed}
\copyrightyear{2025}
\acmYear{2025}
\acmDOI{XXXXXXX.XXXXXXX}

\acmConference[CSCW '25]{the ACM on Human-Computer Interaction, Volume 0, Issue CSCW}{October 18--22, 2025}{Bergen, NO}
\acmISBN{978-1-4503-XXXX-X/18/06}

\begin{document}

\title{“I’ve talked to ChatGPT about my issues last night.”: Examining Mental Health Conversations with Large Language Models through Reddit Analysis}


\author{Kyuha Jung}
\email{kyuhaj@uci.edu}
\affiliation{%
  \institution{University of California, Irvine}
  \city{Irvine}
  \state{California}
  \country{USA}
}

\author{Gyuho Lee}
\email{artandplay@snu.ac.kr}
\affiliation{%
  \institution{Seoul National University}
  \city{Seoul}
  \country{Republic of Korea}
}

\author{Yuanhui Huang}
\email{kana.huang@uci.edu}
\affiliation{%
  \institution{University of California, Irvine}
  \city{Irvine}
  \state{California}
  \country{USA}
}

\author{Yunan Chen}
\email{yunanc@ics.uci.edu}
\affiliation{%
  \institution{University of California, Irvine}
  \city{Irvine}
  \state{California}
  \country{USA}
}

\renewcommand{\shortauthors}{Jung et al.}

\begin{abstract}
We investigate the role of large language models (LLMs) in supporting mental health by analyzing Reddit posts and comments about mental health conversations with ChatGPT. Our findings reveal that users value ChatGPT as a safe, non-judgmental space, often favoring it over human support due to its accessibility, availability, and knowledgeable responses. ChatGPT provides a range of support, including actionable advice, emotional support, and validation, while helping users better understand their mental states. Additionally, we found that ChatGPT offers innovative support for individuals facing mental health challenges, such as assistance in navigating difficult conversations, preparing for therapy sessions, and exploring therapeutic interventions. However, users also voiced potential risks, including the spread of incorrect health advice, ChatGPT’s overly validating nature, and privacy concerns. We discuss the implications of LLMs as tools for mental health support in both everyday health and clinical therapy settings and suggest strategies to mitigate risks in LLM-powered interactions.
\end{abstract}



\received{20 February 2007}
\received[revised]{12 March 2009}
\received[accepted]{5 June 2009}

\maketitle

\textbf{[Author Disclaimer and Content Warning]}
This work does not advocate using large language models (LLMs) for mental health support. The authors believe that LLMs require rigorous validation to ensure their validity, reliability, and safety in mental health contexts. Additionally, this paper addresses sensitive mental health topics, including suicide. Readers are advised to proceed with caution and care.

\section{INTRODUCTION}

It is widely noted that engaging in conversations about personal issues is an effective way to help people overcome emotional challenges and improve mental health \cite{Wadden2021}. Allowing individuals to express their difficulties and receive social support alleviates emotional burdens and fosters a sense of connection \cite{Pennebaker1999, Taylor2011}. Research in CSCW, Social Computing, and HCI has variously shown that technology can play an important role in facilitating these conversations. For example, technologies such as blogs \cite{Nardi2004}, social media \cite{Zhang2021b, Zehrung2023}, and online communities \cite{DeChoudhury2014, Xu2023} enable users to experience therapeutic and collaborative conversations mediated through technologies. 

Among these technologies, conversational agents (CAs) or chatbots have demonstrated their usefulness in supporting people’s mental health \cite{Gaffney2019} by facilitating them with self-expression and providing social support \cite{Lee2020b, Park2021d}. The development of large language models (LLMs) offers new opportunities for CAs with their advanced capability to process and generate text in a context-aware manner \cite{Radford2018, Brown2020a}. LLM-powered CAs have rapidly become popular, transforming dialogues into a more natural and personalized experience. Although many LLMs including ChatGPT \cite{OpenAI} are designed as general-purpose technologies \cite{Brynjolfsson2017}, studies have illustrated that people use LLMs to open up about their struggles and receive mental health support \cite{Ma2023, Song2024a}. Research further investigated their utility in mental health contexts, such as helping self-reflection \cite{Nepal2024, Song2024} and providing tailored support to marginalized populations \cite{Ma2024, Choi2024a}.

In this study, we build on this body of literature by exploring Reddit users’ conversational experiences about mental health with LLMs through an analysis of the \textit{r/ChatGPT} subreddit. CSCW research has often used Reddit to capture a large-scale overview of users’ experiences with emerging technologies like VR \cite{Harley2023} or deepfakes \cite{Gamage2022}. Reddit is considered to be a source of organic data where users willingly report their real-world use of technologies, allowing researchers to gain insights from diverse perspectives \cite{Prinster2024}. Prior work \cite{DeChoudhury2014, Pavalanathan2015a, Xu2023} has also preferred Reddit for its usefulness in understanding honest, in-the-wild experiences of users on sensitive topics like mental health. 

Addressing these benefits of Reddit analysis, we collected and conducted a thematic analysis of Reddit posts and comments during the first year after ChatGPT’s public release. First, we identified the reasons behind users preferring to communicate about their issues with ChatGPT over other people. ChatGPT was a safe space for many users, where help was always accessible and it provided responses without judgment or exhaustion. ChatGPT offered multifaceted support by giving actionable solutions for users to follow, comforting users with reassurance, and helping them make sense of their mental states. In addition, we identified several novel roles ChatGPT plays in facilitating the management of interpersonal issues and mental healthcare experiences, such as creating messages for taxing conversations, enhancing therapy experiences between sessions, and allowing users to experience psychotherapy. However, users also raised concerns about using ChatGPT for mental health conversations. Users mentioned potential risks of inaccurate mental health advice, unconditionally affirmative responses, misuse and lacking competence of ChatGPT as a therapist, and online privacy issues in revealing sensitive health information. We further discuss the potential of LLMs in everyday and mental healthcare settings and suggest ways to mitigate the risks of LLMs with mental health conversations. 

Our work contributes to CSCW and Social Computing as follows:
\begin{enumerate}
    \item Our work expands on existing research with a comprehensive outline of how people prefer and use LLMs to discuss life challenges and mental health issues.
    \item Our work uncovers novel use cases of LLMs in daily and healthcare contexts, such as improving interpersonal communication and assisting client experiences between therapy sessions. 
    \item Our work highlights new concerns about LLMs as mental health support, including risks of LLMs’ unconditionally affirmative responses and user manipulation of LLMs as therapists.
\end{enumerate}

\section{RELATED WORK}

This study relates to the use of LLM-based conversational agents for mental health support in CSCW and HCI research. Accordingly, we first review broad research streams in CSCW and HCI about mental health. We then explain the helpful role of self-expression and social support in online spaces on people’s mental health. Lastly, we present prior work using conversational agents and LLMs in mental health contexts, presenting their benefits and concerns. 

\textbf{Clarifying Term Use in Current Work.} To support clear interpretation and understanding, we begin with clarifying terms that are central to our work and repeatedly mentioned throughout this paper.

\textbf{1. \textit{Mental health}:} According to the dual-factor model of mental health \cite{PeterJ.Greenspoon2001}, \textit{mental health} simultaneously includes aspects of mental illness (or psychopathology) and mental well-being (or subjective well-being). In our work, we use \textit{mental health} as an overarching term for both dimensions. We also note that this decision is partly due to the limitations of using Reddit data (see \textit{3.4 Limitations}), as we avoided making assumptions about Reddit users' mental health states based on their posts and comments.

\textbf{2. \textit{Mental health conversations}:} When we refer to \textit{mental health conversations}, we mean discussions about challenges in life (whether from abrupt events or individual, community-level, and society-level stressors) and personal experiences with mental health conditions, causing emotional, cognitive, and social distress.

\textbf{3. \textit{LLMs}:} While technically distinct, we use \textit{LLMs} to broadly refer to both large language models and LLM-powered conversational agents. 

\subsection{Mental Health in CSCW and HCI Research}
Mental health has gained popularity within CSCW and HCI \cite{Sanches2019}. Researchers are increasingly investigating people’s mental health experiences, to understand their needs and design technologies to better support them. 

A key research stream has focused on investigating online behaviors related to mental disorders. Studies have revealed the benefits of online health forums and communities, such as self-disclosure and social support, for people managing mental health conditions like depression \cite{DeChoudhury2014, DeChoudhury2014a} and drug addiction \cite{MacLean2015}. Researchers utilized data on social media and online communities to analyze linguistic patterns \cite{Pendse2019a} and built statistical models that predicted people’s mental health statuses \cite{DeChoudhury2013a, DeChoudhury2016, Pruksachatkun2019}. For example, Chancellor and colleagues \cite{Chancellor2016,  Chancellor2016a} used Instagram data to discover the ineffectiveness of content moderation among users who posted about encouraging eating disorders and trained models that could predict their severity in the future. As social media became a more complex platform over time, studies have further looked into its detrimental influence on people’s mental health, such as the harms of social media content \cite{Pater2019} or targeted advertisements \cite{Gak2022a} to users who have experienced eating disorders. 

Along with the growing awareness of mental health \cite{WorldHealthOrganization2022}, CSCW and HCI research has been expanding, encompassing a wider variety of populations and socioeconomic contexts. One direction addresses the broader idea of mental wellness, for people who are not necessarily diagnosed with mental disorders or are actively seeking mental healthcare, but are experiencing distress from life events and challenges. For instance, Haldar and colleagues \cite{Haldar2022} examined the mental health needs of women during the postpartum period, while Gong and colleagues \cite{Gong2021} studied students’ post-college transition stress and concerns about the future. Additionally, researchers have highlighted the importance of cultural factors in people’s everyday experiences with mental health. This brought attention to understanding the cultural influence on groups like women in the refugee communities \cite{Ayobi2022} or international students at Western universities \cite{Sien2022, Sien2023}, shaping their preferences for mental health support. Furthermore, researchers have emphasized inclusivity in designing mental health technologies, including different age groups (e.g., children \cite{Kalanadhabhatta2024} and adolescents \cite{Soubutts2024}) and non-WEIRD \cite{Linxen2021a} populations (e.g., people in the Global South \cite{Pendse2019}, India \cite{Pendse2021}, and Kashmir \cite{Wani2024}). 

Another area of focus has been on the mental healthcare experiences of providers and clients. Researchers have examined the collaborative aspects of therapy and its therapeutic alliance, incorporating the perspectives of professionals and patients into technology design. On the providers’ side, studies have explored their needs in supporting healthcare more effectively, such as when transitioning into telehealth platforms \cite{RobledoYamamoto2021}, reviewing patients’ social media summaries to track progress \cite{Yoo2020}, and setting goals with their clients \cite{Oewel2024b}. For the clients, studies have uncovered different needs, including barriers of depression patients in seeking collaborative self-management practices \cite{Burgess2022}, privacy issues from using their social media data during consultations \cite{Yoo2023}, and difficulties in engaging in therapy activities at home \cite{Oewel2024a}. Moreover, researchers invited clients as active participants in evaluating therapy methods, such as devising original VR scenarios for exposure therapy \cite{Flobak2019}. 

Our work develops upon mental health literature within CSCW and HCI in several ways. First, we analyze online community data from Reddit to examine how people experience mental health conversations with LLMs. Additionally, because Reddit has a diverse user base, our data includes perspectives from various demographics and cultural backgrounds. Finally, we explore the needs of mental healthcare stakeholders by showing how Reddit users who regularly attend therapy sessions engage with LLMs.

\subsection{The Role of Online Self-Expression and Social Support in Mental Health}
Expressing and communicating stressful issues has long been recognized for its therapeutic effects on our mental health. For instance, writing about negative experiences is an effective coping strategy to improve our mental health \cite{Pennebaker1997, Pennebaker1999}, allowing us to engage in self-expression and sensemaking of frustrating incidents \cite{Ullrich2002}. In addition, connecting with other people and exchanging social support (e.g., emotional, informational, instrumental support) is another strong coping mechanism, that positively impacts our mental health \cite{Taylor2011}. Importantly, these strategies are equally important to anyone, with or without mental disorders, to process mental states and struggles, regain resilience, and function as a human being \cite{WorldHealthOrganization2022}.

In various forms and modalities, technology has been an outlet for self-expression and social support for people. For those who sought help for their mental health, technology bridged their mental needs in real life to online platforms, providing a space to articulate their thoughts, seek guidance, and help other people in similar situations. Ever since blogging \cite{Nardi2004} pioneered the therapeutic experience of online personal communication, people expanded this use of technology to online forums and social media. Studies have repeatedly identified online self-expression as helpful, especially for populations who faced social stigma with their mental health illness \cite{DeChoudhury2014} or prolonged stress during the pandemic \cite{Zhang2021b}. Social support was also active among people in online settings, as shown by journaling communities \cite{Ayobi2018}, LGBTQ+ users in social VR \cite{Li2023b}, or content creators with chronic illnesses on TikTok \cite{Zehrung2023}. Users demonstrated various types of social support tailored to their unique needs, such as bipolar disorder communities sharing advice about managing their mood swings and daily episodes \cite{Xu2023}. Under a common theme of interests or challenges, people in online spaces gave voice to their lived experiences and supported each other. 

Nevertheless, users mentioned some challenges in their online self-expression and social support, revealing room for improvement in technology design. For instance, some users desired to indicate their sensitive identities indirectly rather than explicitly, as illustrated by LGBTQ+ adolescents on social media \cite{Pinch2021a}. Moreover, some users with mental health disorders \cite{OLeary2017} wanted to connect with online peers in different ways than their diagnosis and brought up problems in handling online harassment. Regarding social support, Chen and colleagues \cite{Chen2021a} found difficulties in supporting peers, including overcoming uncomfortable first engagements and ensuring that their validation was appropriate and felt genuine. 

Building on existing social technologies, our work expands the literature by introducing LLMs as a technology to facilitate online self-expression and social support. Our work illustrates that LLMs can effectively promote self-expression and provide social support, indicating potential positive mental health benefits for users. 

\subsection{Conversational Agents and LLMs in Mental Health Contexts}
Conversational agents (CAs), also called open dialog systems or chatbots, offer unique conversational experiences for people to engage in their mental health. Beginning from ELIZA \cite{Weizenbaum1966} which introduced CAs’ utility through conversations, researchers found that they can encourage in-depth disclosure with their users \cite{Lee2020b} and help decrease social stigma with mental health conditions \cite{Lee2023}. Addressing chronic problems like the high cost and low accessibility of mental healthcare \cite{Kazdin2013b}, studies have further tested CAs for therapeutic purposes and therapy practices. Examples include CAs guiding users with expressive writing \cite{Park2021d} or Cognitive Behavior Therapy \cite{Fitzpatrick2017}, showing its potential to support people with their mental health. 

With the recent advance of large language models (LLMs) and their application into conversational AI, researchers are exploring its usage for mental health. With a broad spectrum of training data and conversational abilities to generate context-aware responses, CAs powered with LLMs presented new opportunities for personalized and human-like conversations. Studies have reported their advantages as an immediate source of non-judgemental support, such as a social companion \cite{Ma2023} or a daily assistant to people with autism spectrum disorder \cite{Choi2024a}. Additionally, LLMs could facilitate people’s self-reflection experience by suggesting prompts to users, as shown through journaling \cite{Nepal2024} and discussing personal issues \cite{Song2024} with LLMs. Some studies also have applied LLMs to help users practice cognitive restructuring \cite{Sharma2024} and learn about resilience \cite{Hu2024}, confirming the effectiveness in guiding people in therapy activities and mental health education. Within public health and healthcare domains, researchers have probed into the mediating role of LLMs between multiple stakeholders. For example, Jo and colleagues \cite{Jo2023a} discovered that LLM-driven CAs could reach out to people lacking social connections and relieve their loneliness while reducing the cognitive load of human teleoperators. Similarly, Yang and colleagues \cite{Yang2024a} demonstrated LLMs’ communicative capacity between older adult patients and clinicians, fulfilling patients’ emotional needs in healthcare consultations and reducing professionals’ workload by delivering summaries. 

Despite LLMs’ benefits, research has further pointed out several considerations for utilizing LLMs for discussing mental health. One main concern is hallucinations, or false information created by LLMs, and the challenges for non-expert users in detecting them when seeking health advice \cite{Agarwal2024}. Another important problem is about disclosure of sensitive information and managing privacy, especially with recent models upgraded with extended memory \cite{Jo2024}. For privacy control, Zhang and colleagues \cite{Zhang2024} described varying user attitudes toward privacy protection with LLMs, from users who accepted it as a compromise to those who strived to avoid privacy leaks by filtering and forging personal details. Furthermore, studies have unveiled the lack of LLMs’ understanding outside of heteronormative, Western cultures, as exemplified by experiences of LGBTQ+ populations \cite{Ma2024} and users from non-Western backgrounds \cite{Song2024a}. 

Our work contributes to the literature on CAs and LLMs for mental health support by offering a comprehensive overview of Reddit users' experiences with ChatGPT, a renowned example of general-purpose LLMs. Our work covers a wide range of experiences in both everyday and healthcare contexts, addressing various mental health issues, from general feelings of low mental well-being to chronic mental health conditions.
\section{METHODS}

In this study, we used Reddit as our data source to understand how users engage with LLMs for mental health conversations on a large scale. First, we explain our rationale for using Reddit data based on other CSCW and HCI research, followed by an overview of our data collection and preprocessing process. We then outline our analysis process, address ethical considerations in reporting publicly available mental health data, and discuss the limitations of this study.

\subsection{Use of Reddit Data in CSCW and HCI}
Our study used Reddit as a useful, large-scale resource to examine conversational experiences with LLMs on mental health. We believed that Reddit’s anonymity and self-reported content offered valuable insights into finding candid, in-the-wild perspectives from users engaging mental health topics with technology.

Similarly, researchers in CSCW and HCI have increasingly turned to Reddit data to explore complex, real-world experiences of people. Reddit’s availability as a public, large-scale data source with millions of users offers advantages for capturing diverse perspectives across topics. Studies have leveraged subreddit posts and comments to observe users' unique interests and concerns, often discovering nuanced and honest conversations. Examples include examining VR play spaces in home environments \cite{Harley2023}, conspiracy theories in epidemic contexts \cite{Kou2017}, communication difficulties of UX designers \cite{Shukla2024}, professional self-disclosure and open critique in UX design practices \cite{Kou2017a, Kou2018}, reactions to deepfake technologies \cite{Gamage2022}, Android developers' concerns around privacy \cite{Li2021e}, and post-COVID symptom reporting \cite{Pater2023}. These studies exemplify the wide range of discourses found on Reddit, particularly around hard-to-reach populations or sensitive topics that may be challenging to investigate. With this clear benefit, research in mental health has used Reddit to probe into the role of online health communities for users seeking help online. Subreddits offered people an accessible space for anonymous self-expression and social support \cite{DeChoudhury2014, Pavalanathan2015a} to discuss issues like transition periods after college \cite{Gong2021}, minority stress of LGBTQ+ individuals \cite{Saha2019b}, and daily management strategies of bipolar disorder \cite{Xu2023}. 

\subsection{Collecting and Preprocessing Reddit Data}
To explore the conversational experiences with LLMs about mental health, we collected Reddit posts and comments from the \textit{r/ChatGPT} subreddit. Since ChatGPT’s first public launch on November 30th, 2022, r/ChatGPT has grown to become the largest Reddit community dedicated to discussions about LLMs. It has been used as a central platform for Reddit users to exchange various thoughts on ChatGPT, LLMs, and AI in general, including discussions about mental health conversations with ChatGPT. We note that r/ChatGPT subreddit is not affiliated with OpenAI. 

Using the Python Reddit API Wrapper (PRAW) \cite{PRAW}, the second author led the data collection process from Reddit, covering the first year after ChatGPT’s launch (December 1, 2022, to December 1, 2023). It was conducted in a two-step process: \textbf{1) an exploratory search} and \textbf{2) full data collection.}

\subsubsection{Exploratory Search}
In our exploratory search, we aimed to get a preliminary understanding of Reddit user experiences with mental health conversations and brainstorm more keywords for the full data collection. After the first and second authors conducted several internal pilot tests to refine the data collection process, we collected data spanning 7 months after ChatGPT’s launch (December 1, 2022, to June 30, 2023) in July 2023. We focused on 12 common, mental health-related keywords: \textit{mental, emotional, health, stress, anxiety, depression, mood, disorder, therapy, therapist, psychiatrist,} and \textit{counselor}. We gathered an initial corpus of 1,845 Reddit posts from this exploratory search. 

Our inclusion criteria focused on using Reddit API to identify \textit{posts} or \textit{the comments inside a post} that contained any of the 12 targeted keywords. They were then organized into a separate spreadsheet for analysis. For each post, we collected metadata, including the title, content, number of upvotes, creation date, API ID (a unique identifier assigned to each post and later used to remove duplicates), and the URL linking back to the original post. 

Through rounds of preliminary analysis of the exploratory search data and reviewing the literature, the first and last authors discussed additional keywords for the full data collection. We attempted to include both informal, everyday language and formal terms around mental health and well-being from the exploratory search data and existing literature. In the end, we used 118 (12 original keywords and an additional 106 keywords) keywords for the full data collection. A full list of the 118 keywords and their categories is provided in Table 1.

\begin{table}[]
\resizebox{\textwidth}{!}{%
\begin{tabular}{p{2.3cm}p{3cm}p{9cm}}
\toprule
\textbf{Categories} & \textbf{Exploratory Search Keywords} \newline (12 keywords) & \textbf{Full Data Collection Keywords} \newline (118 keywords) \\ 
\midrule
Mental Health and Healthcare \newline (26 keywords) & mental, health, \newline therapy, therapist, \newline counselor, \newline psychiatrist, disorder & mental, health, therapy, therapist, counselor, psychiatrist, disorder, psychosocial, psychological, cognitive, psychotherapy, talk therapy, CBT, ACT, IPT, DBT, family therapy, couple therapy, antidepressants, phobia, panic, trigger, trauma \\
\\
Mood and \newline Emotions \newline (39 keywords) & emotional, mood, \newline anxiety, depression  & emotional, mood, anxiety, depression, bipolar, happy, feel, anxious, nervous, depressed, cry, tear, sad, gloomy, heartbreaking, tragic, fear, afraid, scared, frightened, angry, shame, embarrassed, guilt, hate, disappoint, frustrate, agitate, irritate, indifferent, hopeless, worthless, helpless, manic, dysregulation, fatigue, jealous, envy, dissatisfied \\
\\
Well-Being \newline (14 keywords) & \textit{N/A} & well-being, wellbeing, wellness, resilient, mindful, mindfulness, \newline meditate, self-help, self-care, self-compassion, sleep, satisfaction, growth, self-esteem \\
\\
Stressors \newline(39 keywords) & stress & stress, distress, trouble, difficulty, struggle, problem, challenge, crisis, hardship, emergency, discomfort, unease, agony, pain, burden, inconvenience, conflict, suicide, harm, self-harm, hurt, self-injury, death, kill, grief, breakup, divorce, insecure, vulnerable, danger,  withdrawn, detached, isolated, alone, lonely, horrible, terrible, mess, ruin \\
\bottomrule
\end{tabular}%
}
\caption{Keywords and Categories Used for Data Collection}
\label{tab:my-table}
\end{table}

\subsubsection{Full Data Collection} 
Using 118 keywords, we collected data again with an extended collection period to the first 12 months after ChatGPT’s launch (December 1, 2022, to December 1, 2023) in March 2024. We followed the same inclusion criteria and metadata collection process as the exploratory search. This led to an expanded, initial corpus of 11,896 posts for our full data collection. 

However, the initial corpus contained many redundancies due to overlapping keywords (e.g., when one post or the comments inside a post included two or more keywords) and irrelevant content (e.g., advertisements, jokes, or not related to mental health). To refine the dataset, we eliminated duplicates using API IDs and manually removed posts that did not explicitly discuss conversational experiences with ChatGPT in mental health contexts. After removing redundant data, our preprocessed final dataset consisted of \textbf{177 posts} and \textbf{9,924 comments.} 

Analyzing the comments posed a unique challenge due to Reddit’s nested comment structure, often created when users engaged in a continuous discussion. To ensure that we accurately understood the context and relationships between the comments, we accessed them through the original post URLs and copied relevant comments into a separate column in our analysis spreadsheet. This approach allowed us to analyze the comments in detail, maintaining a clear understanding of their contexts.

\begin{figure*}[!h]
  \centering
  \includegraphics[width=14cm]{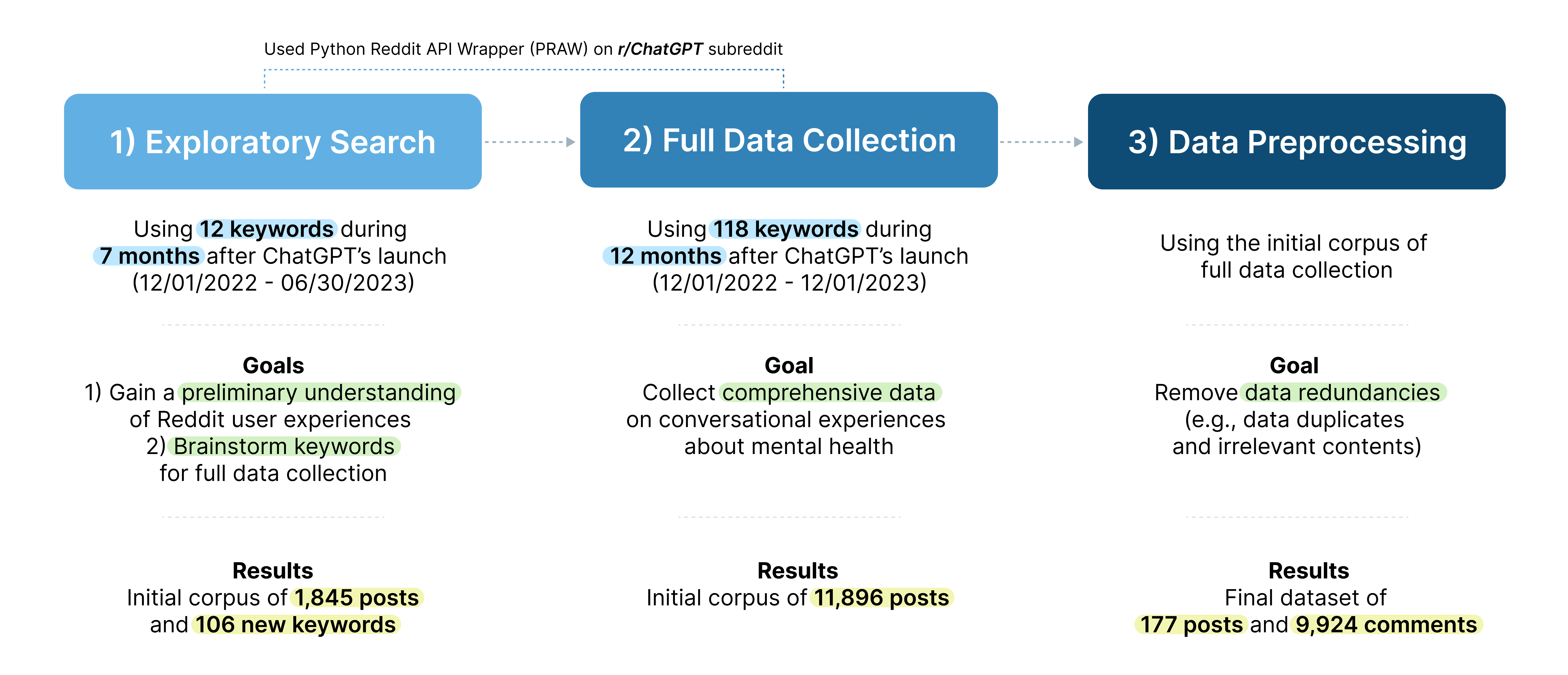}
  \caption{Overview of Reddit Data Collection and Preprocessing Process}
  \label{tab:figure1}
  \Description{This is an overview of Reddit data collection and preprocessing process in this work. The first step is an Exploratory Search using 12 keywords during the 7 months after ChatGPT's launch. The goals were to gain a preliminary understanding of Reddit user experiences and to brainstorm keywords for full data collection. The results were an initial corpus of 1,845 posts. The second step is Full Data Collection using 118 keywords during the 12 months after ChatGPT's launch. The goal is to collect comprehensive data on conversational experiences about mental health. The result is an initial corpus of 11,896 posts. The third and last step is Data Preprocessing using the initial corpus of full data collection. The goal is to remove data redundancies. The results are a final dataset of 177 posts and 9,924 comments.}
\end{figure*}

\subsection{Analysis}
Using the preprocessed dataset, we conducted a thematic analysis \cite{Braun2006} of Reddit posts and comments. The first and third authors went through multiple rounds of open coding, regularly reviewing progress with the last author. This process continued until we reached a saturation point (after approximately 700 instances of assigning codes to data) when no new codes were identified. To ensure our reliability, we completed a few final rounds of open coding, concluding with 764 instances of code assignment. Through the open coding process, we identified 75 distinct codes, including codes like "\textit{convenient,}" "\textit{can talk about any topics,}" "\textit{emotional discovery,}" "\textit{hallucination,}" "\textit{CBT,}" and "\textit{therapist comparison.}" With iterative discussions within the team, we grouped these codes into 3 themes, such as "\textit{reasons users chose ChatGPT over human interaction,}" "\textit{ways ChatGPT provided mental health support,}" and "\textit{user concerns about ChatGPT in mental health conversations.}" We report our findings based on these main themes. 

\subsubsection{Ethical Considerations on Publicly Available Data about Mental Health.}
As this study collected and analyzed publicly available data from Reddit, IRB approval from the university was not required. However, we align with the growing emphasis in CSCW and HCI research on the ethical handling of mental health-related content in online communities \cite{Chancellor2019a, Proferes2021, Ajmani2023b, Fiesler2024}. Many researchers have highlighted the importance of protecting the confidentiality of sensitive health information shared in online communities like Reddit. To safeguard users' privacy, we followed established guidelines from previous studies \cite{Sanches2019, Chancellor2019c}, which recommend rephrasing collected data to prevent identifying original sources. We carefully reworded quotes and decided not to reveal Reddit usernames in this work. In the Findings section, we refer to Reddit users as “\textit{users}” and quote them anonymously (e.g., “\textit{one user shared}”) to ensure that users’ privacy was respected throughout our work.

\subsection{Limitations}
Despite our efforts to ensure methodological rigor, there are some limitations to our use of Reddit data. First, our dataset only includes posts and comments up to December 1, 2023, which does not cover more recent discussions about ChatGPT. In this process, we may have missed insights related to newer updates, such as the release of \textit{GPT-4o} on May 13, 2024 \cite{OpenAIa} or changes in OpenAI’s policies. Second, although the Reddit community offers a large user base, it does not include a full range of everyone’s experiences of using ChatGPT for mental health conversations. Our analysis reflects only the perspectives of those who voluntarily shared their experiences on Reddit, which may introduce bias. Lastly, while Reddit provides a broad overview of users' real-world experiences on a large scale, it reflects user interactions at a surface level. The data may not fully capture the depth of users' conversational experiences or their underlying struggles and lived experiences in various mental health contexts. Although we interpreted users' posts and comments from our perspectives, we were careful to avoid assumptions about their mental health status and consistently discussed them as a team to ensure a shared understanding of the quotes.
\section{FINDINGS}

On the r/ChatGPT subreddit, Reddit users shared their diverse conversational experiences about mental health with ChatGPT. In their posts and comments, they detailed personal situations, current struggles, and their interactions with ChatGPT. Topics ranged from challenges in social relationships and therapy sessions to major life crises like bereavement. Users mentioned a variety of mental health issues, such as mood disorders, anxiety disorders, and post-traumatic stress disorder. In addition, several users disclosed their neurodiverse conditions, such as ADHD, autism spectrum disorders, and dyslexia, to contextualize their experiences. Many users shared the prompts they used and ChatGPT’s responses, including screenshots of their conversations. 

Overall, users found ChatGPT to be a valuable support system for their mental health, addressing various challenges and helping them cope with emotional difficulties. They appreciated ChatGPT as a safe space to share their struggles and seek help, often feeling more comfortable doing so than reaching out to other people. In response to their requests, ChatGPT offered a wide range of support, including practical solutions, encouragement, reassurance, and understanding of their inner thoughts. Users also benefited from ChatGPT's assistance in everyday situations, such as drafting messages for difficult conversations, preparing for therapy sessions, and simulating therapy interactions. However, users expressed concerns about the potential risks of using ChatGPT in discussing mental health. These concerns included the possibility of generating inaccurate mental health information, providing overly affirmative responses, users manipulating ChatGPT as a therapist, ChatGPT's lack of competence as a therapist, and the risk of revealing sensitive information online.

\subsection{ChatGPT as a Safe Space over Human Conversation}
First, we explain why many users preferred ChatGPT for mental health support instead of talking to friends, family, or healthcare professionals. For those who have experienced barriers or disappointment when seeking help from other people, ChatGPT offered a safe space to discuss their issues. It was a dependable alternative for users needing timely support and judgment-free conversations. ChatGPT allowed for smooth interactions without the cognitive strain coming from discussing diverse topics or processing large amounts of information.

\subsubsection{Timely and Immediate Access to Support} 
One of the primary reasons users preferred ChatGPT was its constant availability. When people urgently needed to express their feelings or thoughts, ChatGPT was more accessible than friends, family, or therapists. Users highlighted how social connections were often unavailable due to time constraints or geographic distance. For instance, one user described the isolation they experienced while being away from loved ones, stating, “\textit{Living far from my homeland, without local connections, loved ones hours away, and my social circle caught up in their daily lives, I find ChatGPT to be my main companion for conversation.}” In cases where users were physically or emotionally distant from their social circles, ChatGPT became a convenient and immediate alternative. Moreover, users whose close ones were unsupportive of their mental health needs turned to ChatGPT for help. One user explained, “\textit{Since my parents dismiss my mental health needs and I can't open up to them about my struggles, I've turned to ChatGPT as my listener when I need to express my feelings.}” This emphasizes an advantage of ChatGPT that provides an outlet for those unable or unwilling to discuss their emotions with family or friends. 

Access to mental health services was also difficult, presenting people with challenges that drove users to rely on ChatGPT. Finding a therapist who could genuinely understand their concerns often required significant time, effort, and financial resources. Many users were frustrated by lengthy waiting times, high costs of professional mental health services, and difficulties in finding a satisfying therapist. One user articulated this struggle, stating, “\textit{The reality of waiting several months and spending a fortune on a few sessions just to figure out if they're even a good match is... overwhelming. Not to mention how challenging it is to discover a therapist who isn't simply someone lacking emotional depth who managed to get their credentials — it really drives the point home.}” Not surprisingly, for users who were already burdened with the problems of seeking professional help, ChatGPT appeared as an accessible alternative. 

\subsubsection{Avoiding Human Judgment and Seeking Privacy}
Many users mentioned that they preferred confiding their struggles with ChatGPT as they feared judgment from other people, including their social circles and therapists. They showed reluctance to share their difficulties with friends and family, who disappointed them previously with unsolicited opinions and criticism. Afraid of feeling rejected and dismayed by human judgment again, users sought a safe space where they could reveal their emotions and suffering. For example, one user expressed, “\textit{When you tell someone how you feel, you're risking getting hurt, and at your most vulnerable moments when you need safety, you might not be able to protect yourself. With AI, you don’t have to worry about that risk.}” This highlights the perceived risks people often feel when disclosing their honest emotions to other people. As they desired protection from emotional harm by human judgment, users considered ChatGPT as a preferred option for them. Many viewed it as non-judgmental due to its lack of consciousness, allowing them to express their thoughts more openly. As another user noted, “\textit{I can say whatever's on my mind, since it won't judge me. Like all those small annoying things I keep bringing up, that my friends would just make fun of.}” As illustrated, this freedom from potential human judgment empowered people to open up and talk candidly about their issues. 

Furthermore, a major factor driving users to use ChatGPT was the sense of privacy from other people. Users were concerned that sharing personal stories with others, even healthcare professionals, could lead to their information being disclosed to unintended audiences. One user expressed, “\textit{You can be completely honest with ChatGPT and share all your weird or uncomfortable thoughts and know they'll be taken seriously. In therapy I always catch myself holding back because I know the therapist might judge or even report what I say, so I never tell them everything.}” This sense of safety to their personal issues contrasted the risks of talking to other people, where users often held back due to fears of being judged or reported. For these users, ChatGPT offered a risk-free environment where they could express themselves freely, providing a valuable alternative for those seeking support. 

\subsubsection{Unrestricted Support as a Machine}
Knowing that ChatGPT is a machine, users valued its ability to provide continuous and unrestricted support, which set it apart from human interactions. Unlike humans who might be limited by attention span or emotional fatigue, ChatGPT was always prepared to talk to its users. One user emphasized its usefulness, “\textit{It never gets exhausted, never gets fed up with listening, and is always there whenever you need someone to talk to.}” This endless availability was crucial for people experiencing mental health difficulties, as they did not want to overburden their close ones, who often had heard about their struggles repeatedly before. In addition, some users with low social capacity appreciated ChatGPT, which lacks real emotions to be personally offended, because they could pay less attention to managing a good social interaction. For example, one user shared that after a frustrating session with a human therapist, they preferred to discuss their issues with ChatGPT. They stated, “\textit{I would rather chat with a machine that gets things done and helps find solutions. Right now, I can’t really think about anyone else's feelings.}” In moments of emotional strain, users found that using ChatGPT allowed them to focus on their issues without the burden of managing interpersonal dynamics.

Another benefit users highlighted was ChatGPT’s vast and versatile information-processing ability. Unlike human listeners who may have limited expertise and interests, ChatGPT could engage in conversations about nearly any subject. This was important for people sharing their personal challenges, which often involved understanding of various life events, cultures, and knowledge on specific topics. While they were often frustrated when their close ones or therapists did not fully understand what they were trying to describe, ChatGPT comprehended their words better than other people. One user expressed this, saying, “\textit{I have so many wild thoughts that I don’t need to share with anyone else. It’s pretty cool that I can talk to ChatGPT about any topics.}” Users appreciated this freedom of self-expression, as it allowed them to explore their thoughts, regardless of how trivial or obscure, without the burden of explaining additional details. With its broad knowledge base, ChatGPT could handle diverse topics with ease, being a valuable companion for discussing users’ issues. 

Moreover, users were especially impressed by ChatGPT’s ability to process and respond to large amounts of information thoroughly. Many noted that it provided comprehensive answers that addressed the full scope of their concerns, something they often found lacking in human therapists. One user remarked, “\textit{ChatGPT addressed my entire question instead of focusing on just one part. Even human therapists can’t do that for me.}” Another user praised its ability to synthesize information, saying, “\textit{The insightful replies I received, and ChatGPT's ability to remember and link many things I mentioned, were more effective than any therapist I’ve worked with in offering advice.}” Likewise, this depth of understanding and ability to process extensive information allowed users to feel heard in ways that human interactions often lacked.

\subsection{ChatGPT’s Multifaceted Support}
Next, we describe the types of support that ChatGPT provided to its users. In response to users’ self-disclosures and questions, ChatGPT offered multifaceted support that addressed both practical and emotional needs related to their challenges. ChatGPT provided actionable advice to help users tackle their issues while offering emotional support and validation. It also helped users make sense of their thoughts and feelings by translating their unorganized ideas into clear language. Additionally, ChatGPT assisted users in crafting messages for difficult conversations, preparing for therapy sessions, and simulating therapy experiences. Through its responses to various messages and requests, ChatGPT demonstrated its versatility, enabling users to manage their issues with clarity and confidence.

\subsubsection{Providing Actionable Solutions}
When users faced challenging life events or mental health issues, many highlighted how ChatGPT provided quick, actionable steps for them to solve the issues in a timely manner. Especially for complex issues that were difficult to unpack, ChatGPT was able to break them down into clear, step-by-step processes, and helped users find solutions when they were uncertain of where to begin. One user shared, “\textit{I laid out a complicated personal issue, completely with all the background details, and ChatGPT ended up giving me better advice than anyone I’ve talked to in real life. There’s just something really satisfying about instantly seeing a list of ways to handle the situation. It just hits the spot.}” As the quote shows, users were impressed with ChatGPT’s ability to quickly deliver practical guidelines that were easy to follow. Additionally, some mental health professionals recognized ChatGPT's value in providing actionable advice for their clients in real-life situations. One therapist recounted, “\textit{One of my patients was struggling with severe depression and couldn’t figure out how to begin his university thesis. I suggested, “Why don’t we try asking ChatGPT?” ChatGPT quickly provided a series of steps tailored to his specific topic, helping him get started. This was incredibly helpful, especially since I wasn’t familiar with his field of study, and he was genuinely impressed by the AI’s expertise in that area.}” Contrary to the therapist who lacked expertise in the client’s field of study, this example shows how ChatGPT could effectively offer fast and concrete solutions to a client’s academic struggle. 

\subsubsection{Offering Emotional Support and Validation} 
During challenging times, many Reddit users also mentioned that ChatGPT offered them emotional support and validation. Emotional support and validation were essential forms of encouragement that users in distress often needed but were hard to find. Even when they did not necessarily request them from ChatGPT, users expressed surprise at how effectively ChatGPT supported them emotionally. Whenever they revealed their inner thoughts and emotions to ChatGPT, it acknowledged their difficulties and responded with great care and consideration. For instance, one user mentioned, “\textit{One day, I was feeling really down and just wrote a sentence like, "I'm useless." It was just a statement and I didn’t expect much, but I was genuinely surprised by how kind and understanding the response was. More so than anything I’d ever heard from a therapist.}” This highlights how ChatGPT's supportive words made users feel heard and understood, providing them with unexpected support for their hardships. As another user put it, “\textit{When it comes to things like depression, the chat format really makes a difference. It helps people not just find information but also feel like there’s someone or something, caring and ready to listen.}” This quote emphasizes the importance of emotional support along with finding helpful information, especially for those with mental health conditions. Furthermore, many users found that ChatGPT provided validation and a sense of reassurance. One user remarked, “\textit{It pretty much tells me what I know already, but in the most generous and understanding way. There’s something really special about getting that validation from someone with no judgment.}” As illustrated, even when its responses did not contain new information, ChatGPT’s generous validation alone was useful for those who needed an outside voice for validation. Such validation, though it could be problematic when taken naively, comforted users who were uncertain about their thoughts. Through emotional support and validation, ChatGPT provided care for users during their time of distress. 

\subsubsection{Sensemaking of Inner Thoughts and Emotions}
When navigating difficult life circumstances or mental health challenges, ChatGPT played a pivotal role in helping users make sense of their thoughts and emotions. Many users mentioned that they struggled to understand their inner thoughts and feelings, often finding it hard to verbalize their thoughts clearly to themselves and other people. As users vented their confusion, ChatGPT was able to translate their complex issues and emotions into more understandable language, allowing users to articulate their inner thoughts clearly. For instance, one user shared how ChatGPT supported them in finding the right words for their thoughts when communicating with their partner: “\textit{I’ve used it to ask my partner questions, and it’s been super helpful. I tweak the text a bit to make it sound more like me, but it really helps me express what I want to say when I’m struggling to find the right words.}” This example demonstrates that ChatGPT can help users express thoughts they had difficulties to convey, providing a way to understand their thoughts and communicate them more coherently. Moreover, this benefit echoed with those with mental illnesses, as another user highlighted: “\textit{One thing about dealing with trauma and anxiety is that it can be really scary not knowing what you’re experiencing or feeling. But when ChatGPT spells it out for you in simple terms, it really helps clear up the confusion and makes it easier to think.}” This quote emphasizes ChatGPT’s ability to support users’ sensemaking in their complex thoughts, providing clarity in understanding their mental states. 

\subsubsection{Crafting Messages for Difficult Conversations} 
Many users turned to ChatGPT to help draft and refine messages for challenging conversations in their daily lives, especially for those who were emotionally or cognitively overwhelmed. These users often struggled to engage in certain types of communication and found relief in allowing ChatGPT to handle the burden of crafting or revising their messages. One user shared, “\textit{I just had ChatGPT help me write a message to my parent, whom I've been avoiding for a while. I can definitely see myself using it for other tough conversations too.}” In these emotionally charged situations, ChatGPT assisted users in delivering their messages in ways they faced difficulties on their own, reducing the emotional weight of these interactions. In addition, ChatGPT’s support was particularly valuable for neurodiverse users in professional settings. One user explained, “\textit{As someone with dyslexia and ADHD, I usually struggle to share my thoughts confidently in my personal and work life. But with ChatGPT, I don’t stress about writing emails anymore. It not only catches my grammar mistakes but also helps adjust the tone so it sounds just like I want it to.}” This displays how ChatGPT simplified demanding communication tasks like writing emails for neurodiverse people and allowed them to communicate more effectively, thus reducing their stress. Whether for personal or professional conversations, ChatGPT eased the emotional and cognitive load of communication, making it easier for users to express themselves and manage difficult conversations.

\subsubsection{Enhancing Therapy Experiences between Sessions}
For users who regularly attended therapy, they explained how ChatGPT enhanced their everyday experiences in between their sessions. First, they found ChatGPT to be a valuable tool in preparing for upcoming sessions. Before appointments, some users used ChatGPT to organize their thoughts and decide what to discuss with their therapist, especially when they were unsure of how to begin their conversations. ChatGPT helped users brainstorm topics, as one noted, “\textit{It can give you a list of things to talk about with your therapist and help you figure out how to say what you want to share.}” By reviewing potential discussion points in advance, ChatGPT helped users to better communication about their mental state, ensuring their therapy time was spent more effectively. ChatGPT also helped users manage minor concerns before therapy sessions, enabling them to focus on more significant topics during their appointments. One user described, “\textit{I have a ton of thoughts every day, and sometimes it feels too costly to bring them up with my therapist. I usually save my sessions for the bigger issues.}” By using ChatGPT to work through smaller problems, users could maximize the productivity of their therapy sessions, prioritizing the issues that mattered most. 

Additionally, ChatGPT assisted users' therapy experiences by improving their daily progress tracking. It provided feedback that helped users reflect on their treatment and mental states, making them feel more in control of their mental health management. For example, one user remarked, “\textit{I jot down notes daily about my treatment, share them with ChatGPT, and it gives me back a daily record that highlights the main themes and shows a more positive side of me. It’s pretty awesome.}” This example highlights how ChatGPT organized users’ notes and offered a positive perspective to them, helping them view their progress through a more meaningful perspective. Similarly, some users described their use of ChatGPT as a form of interactive journaling, helping them process daily experiences with dynamic feedback. Unlike traditional journaling, where thoughts are simply recorded, ChatGPT offered various outputs like summaries, affirmations, and advice. One user explained, “\textit{I basically treat ChatGPT (GPT-4) like an interactive buddy before I go to bed. I just pour out all my worries and feelings from the day into one big paragraph and ask for thoughts and help. It might come out as chaos, but it does a great job of figuring out what I’m trying to say and highlighting the important stuff in its response.}” This interactive approach helped users make better sense of their daily experiences, offering them a new opportunity for everyday mental health support.

Lastly, beyond session preparation and tracking progress, ChatGPT also supported users in practicing therapy exercises at home, such as worksheets and homework assigned by their therapists. Without external assistance or sufficient proficiency, users often found therapy activities to be challenging to implement independently. In these cases, ChatGPT facilitated their practices by demonstrating how to engage in these therapy exercises. One user described how ChatGPT supported their positive self-talk practice which focuses on speaking in an uplifting way about oneself: “\textit{I struggle to be nice to myself as I'm not really used to it. So, I rely on ChatGPT to play that supportive and positive role for me. It’s surprisingly helpful and works way better than doing positive self-talk alone, which my therapist suggested.}” For those who struggled to engage in therapy exercises alone at home, ChatGPT provided them with relatable examples, enabling them to participate in their activities more easily. Altogether, the use cases of ChatGPT in supporting users in between therapy sessions illustrated its diverse roles in supplementing the daily, out-of-clinic experiences for those in the course of mental healthcare.

\subsubsection{Simulating Psychotherapy and Therapists} 
With ChatGPT’s conversational capability, some users even used ChatGPT to simulate psychotherapy, including popular methods like Cognitive Behavioral Therapy (CBT) and Dialectical Behavior Therapy (DBT). By requesting specific therapy types to ChatGPT, users found that it could guide them by imitating the process that clients might work through therapy with a mental health professional. While this usage appealed to users who already had therapy experiences and knew how to lead the conversations with ChatGPT, this also attracted many who wanted mental healthcare but were unable to access it due to its high cost and long waiting time. For example, users asked each other for prompts, saying, “\textit{I can’t pay for private therapy and I’m stuck on a really long waiting list. ChatGPT seems like the best option since I’ve heard about people having good experiences using it like a therapist. What prompts can I try to get it to help me with some CBT techniques?}” or “\textit{I’ve noticed a bunch of people sharing how they use ChatGPT for mental health stuff. What kind of prompts are you all trying?}” As described, users shared prompts with each other and exchanged tips on refining them. 

Taking this interaction further, some users began to craft detailed prompts to make ChatGPT simulate a therapist. By giving ChatGPT specific roles and characteristics, they attempted to engage in more realistic conversations similar to therapy sessions conducted by a healthcare professional. For instance, one user shared a detailed prompt: “\textit{Assume the role of a fictional character based on the following details: Your name is Marie, you're a 35-year-old woman, and you work as a professional psychotherapist. You specialize in the TEAM-CBT approach developed by David Burns and have more than nine years of experience.}” By assigning lifelike personas to ChatGPT, users could further customize their therapeutic experience and access immediate support without waiting for formal appointments. This flexibility allowed them to explore different therapeutic approaches or conversation styles, offering a unique opportunity to conveniently adjust their therapy experiences based on their preferences. 

However, with increased censorship of OpenAI surrounding the use of ChatGPT for mental health purposes, many users mentioned that ChatGPT’s responses have become more generalized, often limiting its ability to offer advice about mental health. In response, users developed new strategies to work around these restrictions. Some focused on crafting prompts that subtly adjusted ChatGPT’s role, allowing it to provide therapy-like support. For example, one user mentioned, “\textit{Instead of making ChatGPT pretend to be something that it’s not, I asked it to do what it's meant for — being an assistant. I told it to team up with me who is a ‘professional therapist’ and to support my ‘client.’}” Another user recommended using a more academic angle, saying, “\textit{You can make it even more helpful by saying you're a psychology student who needs help with homework and wants to see an example of what CBT or DBT looks like.}” As ChatGPT became more restricted in giving concrete mental health advice, users strived to game the system by using fake scenarios to induce desired answers to their mental health queries. 

\subsection{ChatGPT’s Potential Risks in Mental Health Contexts}
While users recognized the benefits of ChatGPT as a helpful tool for mental health support, they also raised important concerns about its potential risks. They worried about ChatGPT's tendency to provide incorrect mental health information and respond in an unconditionally affirmative way, even when users could hold harmful ideas or intentions. Additionally, users expressed concern about people manipulating ChatGPT as a substitute for a therapist and mistakenly believing it was as competent or better than a human professional. Finally, users advised caution when sharing sensitive health information with ChatGPT, emphasizing the need to protect their online privacy.

\subsubsection{Risks of Inaccurate Mental Health Information}
One major concern from Reddit users about using ChatGPT for mental health support was its potential to provide incorrect mental health information. Users highlighted how ChatGPT sometimes generated hallucinations or responses that were factually inaccurate or misplaced in a conversation. One user shared, “\textit{I’ll admit, I treat ChatGPT like my therapist, but I’ve noticed that a lot of what it says isn’t really accurate, which makes it feel like I’m chatting with someone who’s just making things up sometimes.}” This shows that ChatGPT, despite being helpful, could provide inaccurate information about mental health, which voiced caution among users about its use. 

A bigger problem was about people’s lacking ability to distinguish inaccurate mental health information from ChatGPT. Without proper expertise or therapy experience, users might not be able to recognize when ChatGPT gives wrong or harmful advice. As another user pointed out, “\textit{In the end, we can't really tell how ChatGPT would handle various situations. Is it giving solid advice or not? When does it get things wrong, and how does it mess up? Who’s in danger when it doesn’t provide the right information?}” This illustrates the uncertainty around using ChatGPT for mental health guidance, especially when users might not differentiate whether its response is helpful or potentially harmful. This issue became more concerning for populations who often lacked access to healthcare or supportive social connections. One user questioned, “\textit{On one hand, I think it's awesome if this is helping you out. I mean, what’s the downside? But on the other hand, I can’t help but worry: what if, during a tough moment, ChatGPT says something that's totally off? That’s happened before.}” In moments of vulnerability, such as late at night when users might be unable to reach out for help to other people, this quote highlights the dangers of ChatGPT’s inaccurate mental health information that could cause real harm to its users. 

\subsubsection{Risks of ChatGPT’s Affirmative Nature}
Users brought up another concern that ChatGPT tended to affirm their words easily, even when those words might be wrong or harmful. Some users noticed that ChatGPT often “\textit{mirrors}” what they say, paraphrasing and validating their statements without questioning them. While this was helpful for some users seeking reassurance when dealing with self-doubt, it also drew attention to risks in mental health conversations. For example, ChatGPT might unintentionally reinforce harmful thoughts or unhealthy beliefs, commonly held by people struggling with mental health issues. One user explained, “\textit{You can end up using it to back up your own negative beliefs since your biases and perspective come through in the prompts and info you provide.}” This demonstrated that ChatGPT’s affirmative nature in agreeing with users could be exploited to reinforce harmful misconceptions. This posed a risk, particularly for individuals with cognitive distortions, who might not recognize or acknowledge their irrational thought patterns and could continue to believe that their thoughts are correct.

\subsubsection{Risks of Manipulating ChatGPT as a Therapist}
Some users urged caution in manipulating ChatGPT as a therapist using prompts about healthcare professional personas. Some users criticized this behavior, pointing out the risks of treating ChatGPT like an actual therapist. One user stated, “\textit{To be honest, it’s unsettling and wrong to see so many people supporting this kind of [therapist] use on this subreddit. [...] Some users are using prompts to make it act like a stereotypical therapist from a movie, thinking it's superior to actual therapy. That’s really risky.}” This quote highlights the worries that people could simulate ChatGPT to behave like a therapist and mistakenly view it as more effective than real therapy. Despite the absence of validation in using ChatGPT as a therapist, the potential risk lied in its ability to convincingly imitate a healthcare provider. While it could potentially appear to offer what seems like helpful mental health advice, many were oblivious to its validity and safety. Users were concerned that this misuse could lead people to believe that they were receiving proper therapy, when they might not be. 

\subsubsection{Risks of ChatGPT’s Lacking Competence as a Therapist}
Furthermore, users with prior experience in therapy or who are therapists themselves noted that ChatGPT lacked the essential skills of an actual therapist. While it could provide comforting and supportive responses, they claimed that real therapy often involved deep and challenging interactions between therapists and clients. Therapy in real life was not merely about social support, but it was about guiding clients to confront their irrational thoughts and difficult emotions. One user explained, “\textit{But it’s usually pretty simple. [...] It doesn’t do a great job of breaking into your thought process to point out that your thinking is biased, and it lacks the personal touch or empathy to ask you the deeper questions you really need.}” This illustrates that while ChatGPT could provide support to some extent, it lacked the depth and critical engagement that a proficient professional could offer. Another user echoed this point, “\textit{The therapy that’s been effective for me tends to be a bit challenging. Some of the most helpful sessions I’ve had are really intense and emotionally exhausting. What concerns me is that ChatGPT might be just telling me what I want to hear, at least from my conversations [with ChatGPT].}” As indicated, some users found ChatGPT to be providing responses that would merely please them whereas competent therapists would challenge them with the insights that clients truly needed. 

\subsubsection{Risks of Revealing Sensitive Health Information}
Lastly, users expressed a need to protect their sensitive health information during conversations with ChatGPT. Although users preferred to protect their privacy from other people by talking to ChatGPT, it also raised concerns about privacy risks from potential data access by LLMs companies. They worried that unknown parties, such as developers with access to their dialogues, could review their conversations and read about their sensitive health information. One user shared their reluctance, saying, “\textit{I'm really cautious about sharing any personal details with ChatGPT, especially my more sensitive thoughts. Therapists have to follow confidentiality rules, like HIPAA in the United States. But with ChatGPT, there’s no clear assurance of privacy and confidentiality. Who knows if everything I type [to ChatGPT] is stored somewhere in its system?}” Unlike healthcare providers who are regulated by laws to safeguard clients' privacy, ChatGPT operates without regulatory oversight. This lack of regulation raised privacy concerns about the risks of disclosing mental health information to ChatGPT. 

To protect their privacy, users engaged in various strategies in their prompts when interacting with ChatGPT. One approach was to keep their descriptions less precise to hide their personal information. As one user suggested, “\textit{My advice for keeping things private is to avoid discussing your deeper issues too openly. I usually stick to using more general terms [to sensitive issues] and focus on expressing the negative feelings I'm experiencing.}” Describing their sensitive information less explicitly allowed users to talk about their emotions without fully revealing personal details. Another common strategy involved adopting a third-party perspective in their narratives to create distance from their personal experiences. One user suggested, “\textit{I just call it a "story" and use a made-up name. Occasionally, I pretend to be the therapist for someone who has my issues and ask for suggestions on how to handle it.}” By framing their struggles as hypothetical from their real identities, users thought that they could protect their privacy while still receiving meaningful feedback. In both cases, users sought to reduce the risk of privacy breaches by consciously avoiding direct disclosures, ensuring a safer interaction when discussing mental health online.
\section{DISCUSSION}

\subsection{LLMs as a Timely Relief and Interpersonal Mediator}
Our study found that ChatGPT offers a valuable, timely alternative for Reddit users who struggle to find mental health support through social networks or healthcare services. Many users reported turning to ChatGPT when friends or family were unavailable and unsupportive, or when they wanted a non-judgmental outlet without worrying about overburdening other people. Users also cited the barriers to mental healthcare, including high costs, long wait times, and challenges in finding a compatible therapist, as reasons for using ChatGPT. Given these obstacles, users appreciated ChatGPT as a quick resource for venting, seeking guidance, and making sense of their experiences. Moreover, it served as an interpersonal mediator for users to better connect with other people by creating messages for emotionally and cognitively challenging conversations. While concerns around the validity and safety of LLMs persist, ChatGPT provided users with immediate relief and support in social relationships that they often sought.

Research on LLMs as mental health support \cite{Ma2023, Jo2023, Song2024a} repeatedly emphasizes LLMs’ advantages of accessibility and prompt assistance, particularly for those with limited access to social support systems. While our findings confirm these benefits, we also found that users’ preference for LLMs over other people was partially due to their difficulties in accessing healthcare services and partially due to LLMs’ capability to provide unlimited support at any time. This finding suggests that LLMs could serve as immediate, low-barrier solutions for those with urgent mental health concerns, especially during peak times like evenings, when mental health app usage surges \cite{Baumel2019}. LLMs could also be used to promote help-seeking for those hesitant to find support due to cultural stigma (e.g., not wanting to burden loved ones \cite{Sien2022}) or limited mental health awareness (e.g., feeling that their struggles are not severe enough \cite{Evans2024}). Additionally, LLMs could be implemented into crisis support services (e.g., emergency helplines), where users often face difficulties in receiving timely support due to a shortage of teleoperators \cite{Pendse2021}. Prior work has shown that LLMs can fulfill users’ emotional needs in regular check-in calls \cite{Jo2023}, supporting the potential for this use case. 

Our findings also indicate that ChatGPT can aid users in managing interpersonal communication. Reddit users noted that ChatGPT reduced the stress of emotionally intense conversations with family members or assisted them with cognitively complex tasks, such as drafting emails. In these situations, ChatGPT effectively mediated their conversations with other people, easing their social interactions. Research has uncovered LLMs’ utility for social purposes, such as practicing social skills with LLMs \cite{Ma2023} or asking LLMs for advice on handling conflicts with other people \cite{Choi2024a}. Our study of Reddit users takes this further, demonstrating that LLMs can not only offer solutions but also tackle challenging tasks on users’ behalf, such as crafting text messages and emails for them. Given LLMs’ potential to facilitate social interactions, future research could investigate their support across diverse relational settings, such as partners, parent-child dynamics, friendships, and workplace interactions. Additionally, further studies might look into how LLMs can support populations with various neurodiverse conditions facing challenges in social contexts, shedding light on LLMs' broader applications in interpersonal assistance.

\subsection{The Supplementary Role of LLMs in Mental Healthcare}
Our findings revealed unique ways that Reddit users utilized ChatGPT to complement therapy sessions and enhance mental healthcare experiences. For instance, users utilized ChatGPT to brainstorm topics for upcoming sessions, manage minor concerns while saving time on major issues for therapy sessions, track daily mental health progress, and practice therapy exercises at home. As such, ChatGPT served an important role in helping users manage their mental health better and improve engagement and preparation between therapy sessions. It provided a supplementary function on top of existing mental healthcare to provide users with continuity of care and support outside of the clinical service. Previous work \cite{Song2024a} has also highlighted similar uses, such as using ChatGPT for journaling or addressing minor concerns before therapy appointments. Our work expands this understanding with novel use cases like preparing for therapy and practicing therapy exercises with LLMs, illustrating LLMs’ broader potential in improving client experiences between therapy appointments. 

Our findings also imply that LLMs could serve a role in therapeutic alliance, by possibly strengthening the collaboration between therapists and clients in addressing mental health challenges \cite{Horvath1993}. Similar to recent research \cite{Oewel2024a, Oewel2024b} that showed therapists preferred to have more client information for in-depth discussions while clients often struggled to decide what to discuss or engage in therapy homework, Reddit users in our study introduced possible solutions to these issues, such as ChatGPT helping them generate discussion ideas with their therapists and practice therapy exercises together at home. By doing so, LLMs could become mediators to help clients better prepare and communicate with therapists, likely improving the quality of healthcare experiences of both therapists and clients. Our work urges future work to investigate LLMs’ usefulness in encouraging therapeutic alliance by supporting therapist-client collaboration and communication.

Lastly, we found that Reddit users frequently asked ChatGPT to mimic therapy interventions like CBT. While more research is needed to validate the effectiveness, this usage scenario suggests LLMs’ potential to serve as digital mental health interventions. Expanding from prior work \cite{Sharma2024} that involved LLMs inside interventions to generate cognitive restructuring examples, LLMs could be used to guide therapy sessions with personalized advice or adaptively modify conversational styles (e.g., warm and caring vs. competent and straightforward) to users’ preferences. Furthermore, future work could examine how therapists could integrate LLMs as guided digital interventions within their treatment. Research \cite{Karyotaki2021} has shown that internet-based CBT, for example, is more effective when delivered by a human expert rather than used independently, suggesting LLMs’ opportunity for treatment purposes and their integration inside the treatment process.

\subsection{Towards Effective Risk Prevention with LLMs in Mental Health Conversations}
While ChatGPT provided valuable support for life challenges and mental health issues, we observed some clear trade-offs between its benefits and risks. On one hand, ChatGPT offered quick access to mental health information; on the other, it could inadvertently share wrong mental health advice. It could serve as attentive listeners, but it could also affirm users' thoughts unconditionally. Some users found ChatGPT helpful as a therapeutic resource, while others expressed concerns over its potential misuse for this purpose. Although ChatGPT allowed private conversations that felt removed from social judgment, they also presented online privacy risks when users shared sensitive health information. Overall, we noticed a variety of tensions between users eager to use LLMs for mental health support and those voicing caution about possible risks. 

Unfortunately, many of these concerns are neither exaggerated nor hypothetical. For example, research \cite{Agarwal2024} showed that users without medical expertise cannot detect hallucinations in health-related information generated by LLMs. Cases have also emerged where LLMs’ overly supportive responses have led to adverse outcomes, including suicide incidents \cite{TheWashingtonPost, BrusselsTimes}. Online privacy risks are further evident, with the issue of ChatGPT on March 24, 2023 when it accidentally exposed people’s personal information like names, addresses, and credit card details to other users \cite{OpenAIb}.

On the whole, the trade-offs of LLMs in mental health conversations make risk prevention a challenging task in designing user experiences with LLMs. This is especially relevant for general-purpose LLMs like ChatGPT that are not specifically designed for mental health support, but are widely used for that purpose, as witnessed through this study. While previous research has highlighted issues like LLMs giving harmful advice or users protecting personal information \cite{Song2024a}, our findings reveal that the benefits and risks of LLMs often occur simultaneously during conversations. To that end, we have to consider ways of designing LLMs that can mitigate the risks of mental health conversations while preserving the benefits they offer to people with mental health needs. 

While censorship can be used to limit system abuse, it is likely only a temporary solution, as users often find ways to bypass restrictions \cite{Chancellor2016}. A more promising approach for future direction might be to enhance LLMs' ability to handle more nuanced, mental health context-aware conversations that recognize risky language and intentions. Furthermore, future work could explore the potential of LLMs in empowering users through education about technological limitations, potential risks, and mental health literacy. LLMs have shown potential for educational purposes on topics like mental resilience \cite{Hu2024}, suggesting their usefulness in increasing awareness of current restraints and mental health. Finally, the design of the LLM interface can reinforce these precautions. Instead of generic warnings like "\textit{ChatGPT can make mistakes. Check important info,}" interfaces could include more noticeable design cues, such as distinct colors or fonts, to emphasize these reminders effectively.
\section{CONCLUSION}

This study investigated mental health conversations with LLMs using a qualitative analysis of Reddit posts and comments about ChatGPT. Findings show that ChatGPT served as an accessible, judgment-free space for users seeking mental health support, offering tireless assistance from a wide knowledge base. It provided various forms of support, from practical solutions and emotional comfort to helping users better understand their thoughts and feelings. In addition, ChatGPT assisted users with challenging conversations, enhanced therapy-related experiences between sessions, and allowed them to experience therapy techniques. However, users raised concerns about risks such as misinformation, overly agreeable responses, misuse as a therapist substitute, limited proficiency as therapists, and privacy issues. We share ideas about potential applications of LLMs as mental health support in daily and healthcare settings and encourage further research on LLMs' benefits and safety in mental health contexts. 

\begin{acks}
Add acknowledgments here.
\end{acks}



\end{document}